\documentclass[11pt]{article}        
\usepackage{times}                   
\usepackage[dvips]{graphicx}

\begin{document}
\title{Neural network prediction of geomagnetic activity:  
a method using local H\"{o}lder exponents}
\author{Z. V\"{o}r\"{o}s and D. Jankovi\v{c}ov\'{a}  \\
 Geomagnetic Observatory Hurbanovo, Geophysical Institute SAS,\\ Slovak Republic\\
  }
\date{}
\maketitle

\begin{abstract}
Local scaling and singularity properties of solar wind and geomagnetic
time series were analysed using H\"{o}lder exponents $\alpha$. It was 
shown that in analysed cases due to multifractality of fluctuations $\alpha$
changes from point to point. We argued there exists a peculiar interplay 
between regularity / irregularity and amplitude characteristics of 
fluctuations which could be exploited for improvement of predictions 
of geomagnetic activity.
To this end layered backpropagation artificial neural network model 
with feedback connection was used for the study of 
the solar wind - magnetosphere coupling and prediction of geomagnetic $D_{st}$
index. The solar wind input was taken from principal component analysis of 
interplanetary magnetic field, proton density and bulk velocity. Superior 
network performance was achieved in cases when the information on local H\"{o}lder 
exponents was added to the input layer.
\end{abstract}

\section{Introduction}
\label{sec:intro}
One of the goals of solar-terrestrial physics is to predict the response 
of magneto\-sphere-ionosphere system to highly variable conditions in the 
solar wind (SW). The question of solar wind-magnetosphere coupling (SWMC) 
can be studied by means of input-output modelling. Linear input-output 
techniques (or linear prediction filtering) describe the SWMC by a 
linear moving-average (MA) filter assuming that the convolution of a 
time-invariant transfer function (TF), with an earlier SW input can predict 
the magnetospheric output represented by time series of geomagnetic indices 
\cite{iye79, bar85, mcp88}. The TF characterizes the magnetospheric response and can be estimated
directly from data provided that a sufficiently large number of input-output
pairs is available. In fact, \cite{bar85} using the $vB_{z} - AL$ input-output data 
($v$ - solar wind velocity, $B_{z}$ - interplanetary magnetic field $N-S$
component, $AL$ - auroral zone geomagnetic index) have shown that the linear 
MA filters can identify two different regimes in which SW energy is dissipated 
within the magnetosphere (directly driven and loading-unloading regimes).
At the same time, the best linear MA filters do not predict the geomagnetic 
output precisely, unless strongly varying filter parameters are considered 
in each case of activity level separately \cite{bla94}. Different levels of geomagnetic 
activity and the nonlinearity of the SWMC were then treated by nonlinear MA 
filters \cite{pri94, vas95} using the assumption that the geomagnetic activity is a nonlinear 
function of the solar wind input. Actually, local linear (that is nonlinear)
MA filters were used, which represent a linear approximation of the nonlinear 
system. Nonlinear MA filters proved to be better predictors of geomagnetic 
response as the linear ones, but the internal dynamics of the magnetosphere and the 
additional influence of it on the geomagnetic response itself (a feedback) 
was more explicitly considered within the frame of state-input space models \cite{vas95}.
Here the prediction of magnetospheric states is made within a common input 
(solar wind) - output (geomagnetic data) phase space and the local linear 
(nonlinear) approximation is given by an evolution of the nearest 
neighbours of a phase space point. \cite{vas95} found that in comparison with 
linear state-input models (global aproximation) the nonlinear state-input 
models (local approximation based on nearest neighbours) give better predictions
of geomagnetic activity.

An alternative to the above MA filters is represented by artificial neural 
networks (ANN) which are global nonlinear functions. 
Elman recurrent 
ANN was used by \cite{mun00} to model SW forcing of the westward 
auroral electroject and the storm-time ring current.
In predicting geomagnetic
activity their performance was similar to that of linear filters \cite{her93}. 
Significantly better performance was achieved by gated ANNs that accounted for 
different levels of activity. \cite{wei99} used three individual ANNs for modelling 
low, 
medium and high $vB_{z}, AL$ activity levels using data from database of 
\cite{bar85}.
The outputs of these ANNs together with past geomagnetic outputs were used to 
train the gate network. It was shown by \cite{wei99} and \cite{wei00} that the gated architecture give 
significantly better predictions as the ungated one or the ARMA system reported 
by \cite{her93}. Obviously, the gated ANN architecture resembles the state-input space 
model of \cite{vas95} giving account for changing activity levels. Local linear filters 
can be calculated in a neighbour of any point in state-input space, the gated 
ANN, however, uses only three levels of activity. 

In this paper we propose a method which allows to consider the changing level 
of SW fluctuations. Instead of building a more structured gated ANN 
architecture we use the extra information on local scaling characteristics of 
properly introduced measure which can be estimated directly from a time series.
Multifractals exhibit time-dependent scaling laws and hence allow a description 
of irregular phenomena that are localized in time.
Multifractal scaling characteristics of geomagnetic fluctuations were studied 
by \cite{cons96} and \cite{voro00}.
\cite{jan1a} using multilayer feed-forward ANN have shown that the 
information on multifractal characteristics of geomagnetic data put to the input 
enhanced the performance of their ANN in 
reconstructing AE-index time series from geomagnetic observatory data. 
The inclusion of multifractality, however, somewhat amplified the noise 
component in this case.
We expect that the inclusion of the scaling characteristics of solar wind and
geomagnetic 
fluctuations to the ANN modelling of SWMC offers a way for considering 
essential local 
information on rapid changes, irregularities and intermittence not considered 
enough hitherto. Intermittence of SW and geomagnetic fluctuations was not built 
into nonlinear filter or ANN models. Notwithstanding that SW fluctuations proved 
to be strongly intermittent \cite{burl91, carb94, mars96, tu96, brun99} 
and also both nonlinear magnetotail theories 
\cite{chan99, chap98, klim00} and experimental works 
\cite{cons96, boro97, cons98, conl99, voro00, kovz01, watk01} 
predict or confirm the presence of scalings, 
multifractality and intermittence within the magnetosphere. Though there exist 
competing theoretical concepts regarding the underlying physical mechanisms which 
may or may not produce the observed scalings \cite{frea00, ant01} 
these considerations have no 
effect on our analysis. We simply ask what are the scaling characteristics of 
fluctuations and how can this information improve our ability to predict 
geomagnetic activity using ANNs.

\section{Data analysis methods}
\label{sec:dat}

\subsection{Local scaling characteristics: the H\"{o}lder exponents}

\label{subsec:des}

We consider the accumulated amount of signal energy within a window 
$W: (t_{i} - W, t_{i})$. The signal energy $E$ within a window 
$W$ is computed as a sum of the squared amplitudes of time 
series through
\begin{equation}
E(t_{i})=\sum_{i-W}^{i} X^{2}(t_{i}); \:\:\:\:\:\: i=1, 2, \dots ,N
\end{equation}
and
\begin{equation}
\sum^{N}_{i=1} X^{2}(t_{i})=1
\end{equation}
where $X(t_{i})$ represents a time series, $N$ is the total number 
of data points. 
The distribution of $E$ in time 
is considered to be a measure which may also appear as singular.
Mathematically, a measure can be characterized by its density. 
An 
erratic behaviour appears in the absence of a density for a singular measure.
Generally, 
singular distributions can be characterized locally by the so-called 
singularity or H\"{o}lder exponents $\alpha$ \cite{hal86, muz94, vehe98}. 
Loosely speaking, 
the exponent $\alpha$ quantify the degree of regularity or irregularity
(singularity) 
in a distribution or a function in a point $t_{i}$. Usually, the measure 
$E(t_{i}, W)$ within a window $W$  scales as $W^{\alpha}$. Therefore, $\alpha$
can be estimated by a regression method using 
\begin{equation}
\alpha(t_{i}, W) = \frac{log E(t_{i},W)}{log W}
\end{equation}
taking different window lengths $W$. 
For a monofractal $\alpha(t_{i})=const$ for all $t_{i}$, while in a case of 
multifractal measure (non-uniform distribution) $\alpha$ changes from point 
to point (non-stationarity). For instance, fractional Brownian motion or continuous 
It\^{o} processes represent self-affine fluctuations governed by a single H\"{o}lder 
exponent. 
The global distribution of singularity exponents $\alpha$ for geomagnetic 
fluctuations was studied by \cite{cons96} and \cite{voro00}. 
It was shown that on the time 
scale of substorms and storms geomagnetic 
fluctuations seem to be analogous to the 
simple multiplicative $p$-model 
which describes energy cascade processes in 
turbulent flows. 
This model explains how a specific energy flux introduced on 
large scales to a flow can lead to non-homogeneous, intermittent energy 
distributions on small scales.
On this basis we expect that in case of 
homogeneous energy transfer rate 
between scales with no intermittency effects, 
the above defined distribution 
will be stationary and 
$\alpha(t_{i}) \sim 1$ for all $t_{i}$.
Otherwise, $\alpha(t_{i}) < 1$ indicate 
irregularities, sharp variations around $t_{i}$, while $\alpha(t_{i}) > 1$
is found in regions where events are more regular \cite{rie97}. In case of 
multifractal processes $\alpha$ changes from point to point, which 
usually makes difficult the numerical estimation of $\alpha$'s. A number 
of papers deals with this question \cite{muz94, jaf96, mal92, vehe98}. 
Though the H\"{o}lder exponents 
do not characterize the local regularity properties of a signal completely 
\cite{gui98}, we are going to use the simple relation (3) to show that even a 
rough estimation of local scaling characteristics of the signal may enhance 
the performance of ANNs.
 We note that a running numerical estimate of
$\alpha$ may fluctuate sharply for other, from multifractality different,
nonstationary processes.

\subsection{ANN description}

\label{subsec:ANN}

A layered backpropagation ANN model \cite{rum86, kro96} 
with feedback connection from 
output layer to input layer was constructed. The output-input layer 
connection makes the output history to be an ordinary input unit in training 
process.
The output of the model can be expressed in the form 
\begin{eqnarray}
y(t+\Delta \tau) = F ( \sum_{k=1}^{Q} w_{k} f_{k} (\sum_{j=0}^{T} v_{jk} I^{(1)}_{j}(t - j \Delta \tau) + \sum_{j=0}^{T} v_{jk} I^{(2)}_{j}(t - j \Delta \tau)
+ \nonumber \\ \sum_{i=0}^{T} u_{ik} y(t - i \Delta \tau) + v_{0}) + w_{0})
\end{eqnarray}
where 
$y$ denotes the $D_{st}$ time series;
the two inputs equal $I^{(1)} \equiv Pc1$ and $I^{(2)} \equiv Pc2$; 
$T$ the history; 
$\Delta \tau$ the time resolution ($\Delta \tau = 1$ h); 
$u_{ik}, v_{jk}$ the weights between input and hidden layers; 
$w_{k}$ the weights between hidden and output layers; 
$v_{0}, w_{0}$ the biases of the layers;
$Q$ the number 
of hidden units; 
$F$ and $f_{k}$ the nonlinear activation function. 
In our model $f_{k}$ are the hyperbolic tangent 
and $F$ the linear activation functions and $Q = 6$.  
The performance of the 
ANN model was evaluated through root mean squared error ($RMSE$) and correlation 
coefficient ($\rho$)
\begin{equation}
RMSE = \sqrt((\sum_{i=1}^{N}(y_{i}^{out} - y_{i}^{pred})^{2})/N)
\end{equation}
\begin{equation}
\rho = \frac{\sum_{i=1}^{N} (y_{i}^{out} - \bar y^{out}) (y_{i}^{pred} - \bar y^{pred})}{\sigma_{y^{out}} \sigma_{y^{pred}}}
\end{equation}
where $y^{out}$ denotes an actual output, $\bar y^{out}$ its mean value 
and $y^{pred}$ a one-step ahead prediction of ANN, $\bar y^{out}$ its mean value;
$N$ is their length; $\sigma_{y^{out}}$ and $\sigma_{y^{pred}}$ are the standard deviations of 
$y^{out}$ and $y^{pred}$.

\section{Data analysis}
\label{sec:dat}

In this paper we are going to predict the $D_{st}$ index one hour in advance 
using the layered backpropagation ANN model with feedback connection.
Prior to that, we show several examples which demonstrate that the H\"{o}lder 
exponents estimated by Equation 3 provide local characteristics of the analysed 
time series sensitive enough to capture the necessary information on the 
abrupt changes and activity levels.

Figure 1a shows interplanetary magnetic field (IMF) variations registered by 
the ACE satellite which is continuously monitoring the SW at the $L_{1}$ 
Earth-Sun Lagrange point. 
The time resolution is 16 {\it{s}} and 5 hours of data is shown from January 14, 
1998, 05:20 UT. This is a time period of very low activity level with mean 
value of IMF ACE $B$ fluctuations of 3 nT. The H\"{o}lder exponents estimated 
within variable window length $W \in (16, 16*160)$ {\it{s}} at each point are depicted 
in Figure 1b. It is visible that $\alpha$ fluctuates around its mean value
$\bar \alpha \sim 1$, which means that the measure is almost uniformly 
distributed. The energy content of the signal $E$, and its scaling with 
window length, that is $\sim W^{\alpha}$, is shown in a log-log plot in Figure 1c.

In contrast with Figure 1, Figure 2 shows a more disturbed period of IMF ACE $B$ 
variations from March 31, 2001 from 00:00 to 05:00 UT. The mean value of $B$ is 
43 nT. Large departures from $\bar \alpha = 1$ are present (Figure 1b), mainly 
within time periods of enhanced fluctuations. These periods are characterized by 
sudden increase of regularity ($\alpha > \bar \alpha$) followed by periods of 
low regularity ($\alpha < \bar \alpha$) or vice-versa.

In fact, $\alpha$ appears to be a sensitive indicator of fluctuations which 
may occur during periods of enhanced IMF $B$ amplitudes, however, when the 
fluctuations cease, the values of $\alpha$ return to $\bar \alpha \sim 1$, 
independently on the actual amplitudes. A good example of it is visible within 
the time interval $t \in (2600, 5000)$ {\it{s}} in Figures 2 a, b, where $B > 50$ nT
and $\alpha \sim 1$. Moreover, the local fluctuations of $\alpha$ around 
$\bar \alpha$ seem to be larger when the gradient of $B$ increases, but it is not 
always valid (not shown). There is also a clear difference between the scalings 
in Figure 1c and 2c.

We conclude that, besides the amplitude of magnetic field variations, the local 
scaling properties of signal described by H\"{o}lder exponents $\alpha$ 
(Equation 3) may represent an essential piece of information the consideration 
of which would allow a better prediction of future geomagnetic activity.

Other examples of longer period  data sets 
(from March 19 to April 25, 2001) are
depicted in Figure 3. This time, IMF $B_{z}$ from ACE satellite and the 
$D_{st}$ index are considered with time resolution of 1 hour. The thick 
line in Figure 3a corresponding to $B_{z} = - 10$ nT highlights periods 
of enhanced SWMC. 

\cite{gonz87} have shown that the interplanetary causes of 
intense magnetic storms ($D_{st} < -100$ nT) are long duration ($> 3$ h)
large and negative ($< -10$ nT) $B_{z}$ events associated with 
interplanetary duskward electric fields $> 5 [mVm^{-1}]$. Comparison of 
Figures 3a, d shows an agreement with the above criteria, that is, long 
duration negative IMF $B_{z}$ events occur together with intense magnetic 
storms. Horizonthal thick line corresponds to the limit of 
$D_{st} = - 100$ nT in Figure 3d. Figure 3b shows the normalized measure
$E$ and the estimated H\"{o}lder exponents are in Figure 3c. Approximately 
the same behaviour is visible as previously (Figure 2), which may be even 
better visualised by drawing 3D plots of time,  IMF $B_{z}$ or $D_{st}$ 
index and the corresponding H\"{o}lder exponents as in Figures 4a, b. In 
both cases when the above mentioned physical limits of amplitudes 
($B_{z} < - 10$ nT and $D_{st} < - 100$ nT) are crossed, the 
H\"{o}lder exponents have their local minima, $\alpha < \bar \alpha$,
indicating sharp irregular variations. Intense 
magnetic storms ($D_{st} \leq - 100$ nT and $\alpha < \bar \alpha$) 
are usually preceeded by 
sudden increases of $\alpha \gg \bar \alpha$, that is, by short 
periods of increased regularity (Figure 4b). The same effect is present 
in $B_{z}$ time series (Figure 4a), though, except the large event around 
$\sim$ 300 hours, it is less visible. 

We expect that precisely the 
interplay between regularity / irregularity and amplitude characteristics 
should be learnt by ANNs to achieve superior performance.
The simplest way to realize that is to add, besides the amplitudes of 
the analysed variables, the corresponding series of H\"{o}lder 
exponents to the ANN input.
The following ACE SW parameters with $\Delta \tau = 1$ hour 
time resolution were used:
$B_{x}$, $B_{y}$, $B_{z}$, $\vert \bf B \vert$, $n$, $v$. The time evolution of 1 hour
$D_{st}$ index from January 1 to July 28, 2001 was considered. The time series of SW
parameters were preprocessed using principal component ($Pc$) analysis
\cite{gna77, rey96}. The linear combinations of normalized SW parameters, 
their derivatives 
and combinations: $B_{x}$, $B_{y}$, $B_{z}$, $\vert \bf B \vert$, $n$, $v$,
$nv$, $n \vert \bf B \vert$, $v \vert \bf B \vert$, 
$vn \vert \bf B \vert$, $dB_{x}/dt$, 
$dB_{y}/dt$, $dB_{z}/dt$, $d \vert \bf B \vert/dt$,  
$dv/dt$, $dn/dt$
were used for the calculation of the $Pc$'s. It was shown by 
\cite{jan1b},
that for the considered set of SW parameters, most of the variance of SW 
fluctuations is controlled by the first two components.
In this paper we use $Pc1$ and $Pc2$ as SW input time series.

%\begin{eqnarray*}
%Pc2 = +0.21B_{x} + 0.15B_{y} - 0.25B_{z} + 0.41 \vert B \vert
%- 0.13n + 0.46v - 0.22nv - 0.17n \vert B \vert + 0.31v \vert B \vert \\
%- 0.18vn \vert B \vert - 0.16\frac{dB_{x}}{dt}
%- 0.18\frac{dB_{y}}{dt} + 0.04\frac{dB_{z}}{dt} + 0.09\frac{d \vert B \vert}{dt} 
%- 0.09\frac{dv}{dt} + 0.16\frac{dn}{dt}
%\end{eqnarray*}

The local scaling characteristics of the principal components are described
in the same way as of the other SW parameters.
The time interval under study was divided into two subsets. The first 
one (part A in Figure 5) from January 1 to March 14, 2001 was used for 
ANN training while the second one (part B in Figure 5) from March 15 
to July 28, 2001 represented independent set for prediction, not included 
in ANN training process. 
The influence of inclusion of local H\"{o}lder exponents on ANN performance 
was tested for a set of values of history $T$ and window length $W$, 
whilst $T = W$.
In all cases analysed here a feedback consisting of past $T$ values of $D_{st}$
index was set. Figure 6 shows the dependence of correlation coefficient 
$\rho$ (Equation 6) in three different cases:
1.) H\"{o}lder exponents $\alpha$ are not considered on input at all - only 
$Pc1$, $Pc2$ and the $D_{st}$ feedback with history $T$  
(indicated by a continuous line); 
2.) H\"{o}lder exponents of $Pc1$ and $Pc2$ vectors are added as input 
(marked by "$*$");
3.) as in case 2.), but H\"{o}lder exponents describing the local scaling 
properties of past $D_{st}$ values are also added as an extra input
(depicted by "o").
The effect of the inclusion of H\"{o}lder exponents is evident mainly in the 
superior performance of ANNs in case 3. The correlation coeficient 
$\rho$ achieves its maximum $\rho_{max}=0.99$ at $W = T = 2$ h and
decreases with increasing $T$ and $W$. At the same time ANN performance 
is practically unchanged in cases 1 and 2 when $T$ and $W$ increase.
We mention that without the $D_{st}$ feedback $\rho$ slowly increases with $T$ 
\cite{jan1b}. As it can be seen, the consideration of scaling properties of $Pc1$ 
and $Pc2$ SW data enhances a little the performance level of ANN,
but a real improvement is achieved when the singularity or regularity 
properties of geomagnetic fluctuations are taken into account, too (case 3).
It seems to confirm our expectation that the information on local scaling 
properies of signals put to the input layer allows to learn input-output
relations better accounting for changing activity levels more effectively. 
The analysis of $RMSE$ (Equation 5) leads to the same conclusion.
For demonstration 1 hour ahead predictions of an intense geomagnetic 
storm are shown in Figure 7a. Two methods are compared (Figure 7b, c): 
case 1 as defined above, 
when the $D_{st}$ index is predicted without H\"{o}lder exponents and case 3, 
with the information on $\alpha$'s ($Pc1$, $Pc2$ and $D_{st}$) added to the 
input layer (the cases 1 and 2 are similar). 
Easy to recognize that the method using $\alpha$'s (case 3) 
allows to predict almost all the variance in the data with $\rho = 0.99$ 
and $RMSE = 2$ nT having $T = W = 2$ h. At the same time $\rho = 0.93$,
$RMSE = 7$ nT for $T = W = 2$ h (Figure 6 in case without H\"{o}lder exponents). 
In comparison, \cite{wul96} 
have exploited Elman recurrent ANNs to predict the $D_{st}$ index 1 hour 
ahead only from SW data. They achieved $\rho = 0.91$ and $RMSE = 16$ nT.

\section{Conclusions}
\label{sec:con}

We presented a prediction technique which uses the extra information on 
local scaling exponents to improve the performance of a layered ANN with 
feedback. 

It was demonstrated that the H\"{o}lder exponents $\alpha$ are time 
dependent and change from point to point exhibiting large deviations from 
the mean value $\bar \alpha = 1 $, mainly during enhanced activity levels 
of fluctuations. A peculiar interplay between regularity / irregularity features 
(described by $\alpha$) and amplitude characteristics of disturbances was 
found and demonstrated on examples of SW and geomagnetic data. ANN performance was 
significantly improved by putting the H\"{o}lder exponent time series of 
corresponding SW and geomagnetic past data to the input layer yielding 
the least $RMSE$ error of 2 nT for short history $T = 2$ h and window 
length $W = 2$ h. The results obtained without 
H\"{o}lder exponents were the worst
($\rho \sim 0.93, RMSE \sim 7$ nT). Only a small improvement if any was achieved 
when the H\"{o}lder exponents of SW $Pc1$ and $Pc2$ were added 
only ($\rho \sim 0.94, RMSE \sim 6$ nT). 
It means that to understand and model better 
the magnetospheric response, in addition to SW input and geomagnetic history 
(feedback), the scaling and irregularity / regularity features of magnetospheric 
fluctuations should also be taken into account. It is not an 
unexpected result, 
however, because recent nonlinear theories on SWMC or magnetotail dynamics involve 
or predict the appearance of scalings, irregularities (singularities) 
and turbulence \cite{gale86, chan99, chap99, klim00}.
To fully exploit this approach on experimental basis, further investigations of 
scalings and singularity features of fluctuations in different inner and outer 
regions of the magnetosphere will be necessary.
\\

Acknowledgements \\
The authors wish to acknowledge valuable discussions with P. Kov\'{a}cs, 
D. Vassiliadis and N. Watkins. $D_{st}$ index 
from WDC Kyoto are gratefully acknowledged. 
We are grateful to N. Ness (Bartol Research Institute) and D.J. McComas 
(Los Alamos National Laboratory) for making the ACE data available. 
This work was supported by VEGA grant 2/6040.

{}
\pagebreak

\begin{figure}[tb]
\centerline{
\includegraphics[width=3.2in]{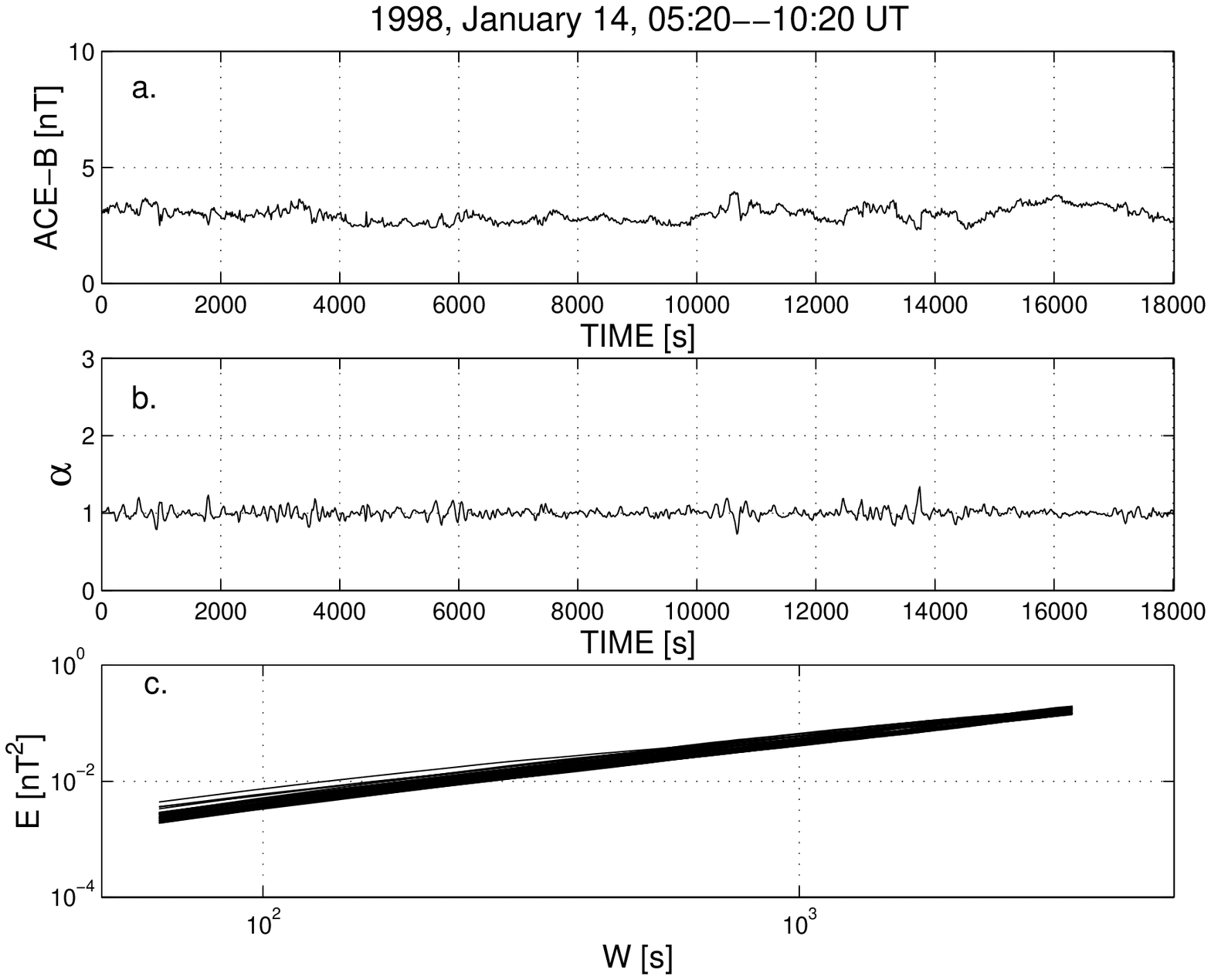}
}
\caption{Period of low activity level;
  {\bf a}. interplanetary magnetic field from ACE satellite (time resolution 16 [s]);
  {\bf b}. the estimated time series of H\"{o}lder exponents $\alpha$
  {\bf c}. the energy content of the signal versus window length $W$.}
\end{figure}

\begin{figure}[tb]
\centerline{
\includegraphics[width=3.2in]{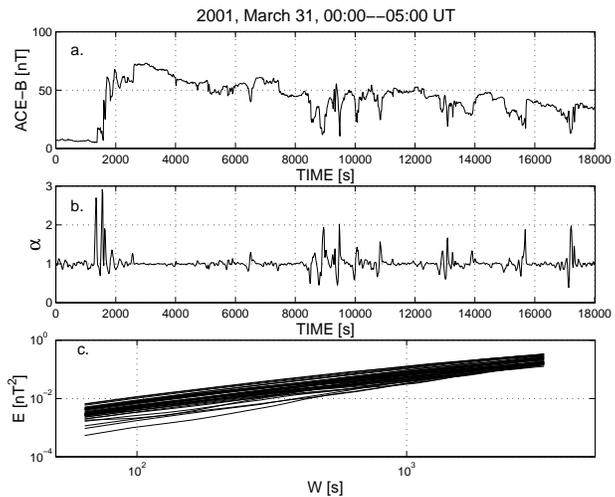}
}
\caption{ Period of high activity level;
  {\bf a, b, c} - same as in Figure 1.}
\end{figure}

\begin{figure}[tb]
\centerline{
\includegraphics[width=3.2in]{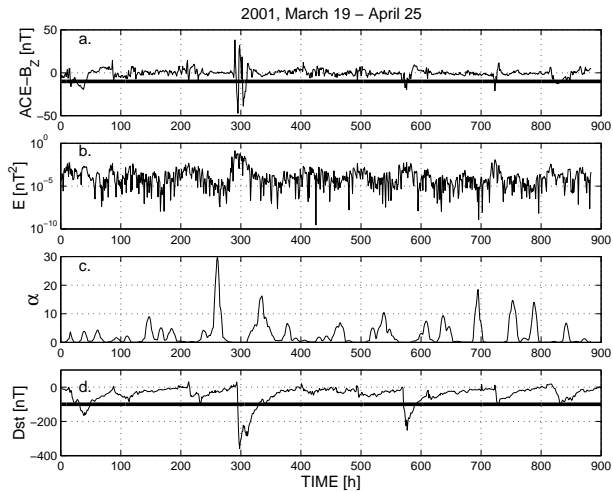}
}
\caption{{\bf a}. Interplanetary magnetic field $B_{z}$ component (time resolution 1 hour;
  {\bf b}. the corresponding energy content $E$;
  {\bf c}. the H\"{o}lder exponents;
  {\bf d}. geomagnetic $D_{st}$ index.}
\end{figure}

\begin{figure}[tb]
\centerline{
\includegraphics[width=3.2in]{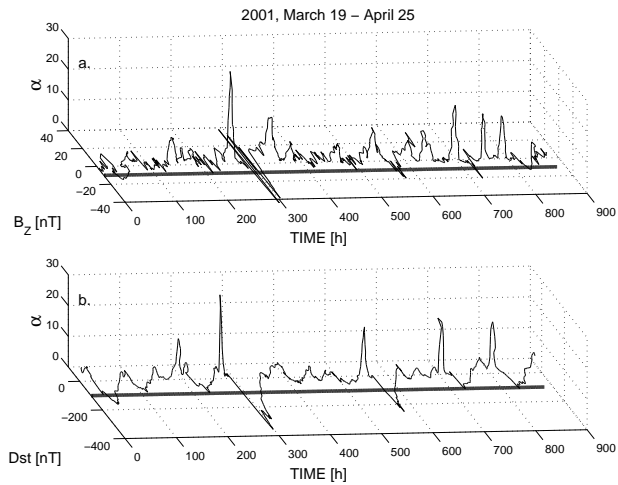}
}
\caption{The interplay between regularity / irregularity and amplitude characteristics;
  {\bf a}. interplanetary magnetic field $B_{z}$;
  {\bf b}. geomagnetic $D_{st}$ index.}
\end{figure}

\begin{figure}[tb]
\centerline{
\includegraphics[width=3.2in]{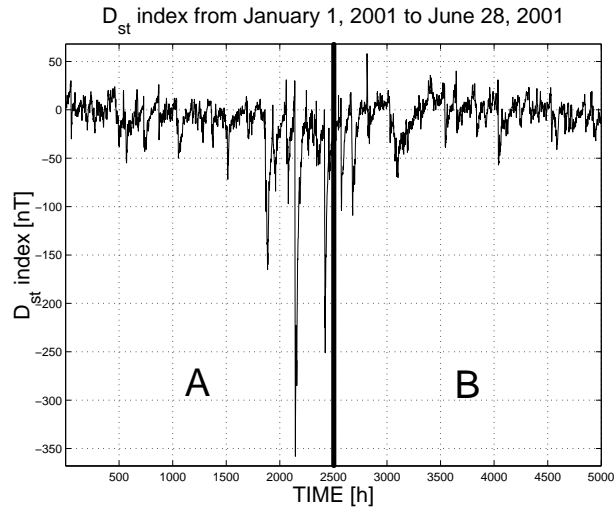}
}
\caption{$D_{st}$ time series from January 1, 2001 to July 28, 
           2001
           used for ANN analysis (A- the period for training process; 
	   B- independent 
	   set for prediction; thick vertical line divides A and B).}
\end{figure}

\begin{figure}[tb]
\centerline{
\includegraphics[width=3.2in]{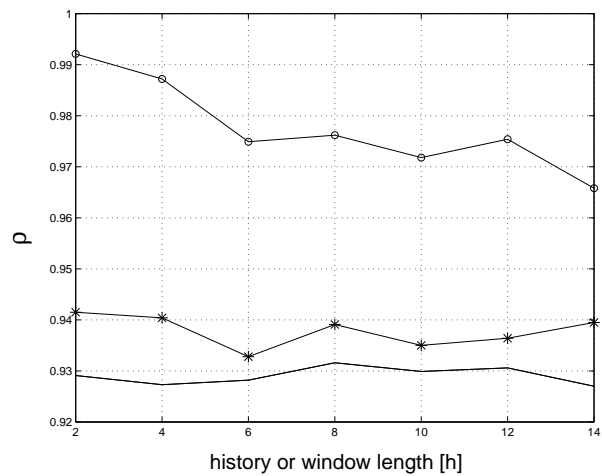}
}
\caption{Correlation coefficient dependence on history $T$ 
           or window length $W$ ($T$ = $W$); -- - without H\"{o}lder   exponents; 
             $*$ - H\"{o}lder exponents for vectors of $Pc1$
	   and $Pc2$ inputs considered; o - H\"{o}lder exponents for inputs 
	   $Pc1$ and $Pc2$ and for past $D_{st}$  index considered.}
\end{figure}

\begin{figure}[tb]
\centerline{
\includegraphics[width=3.2in]{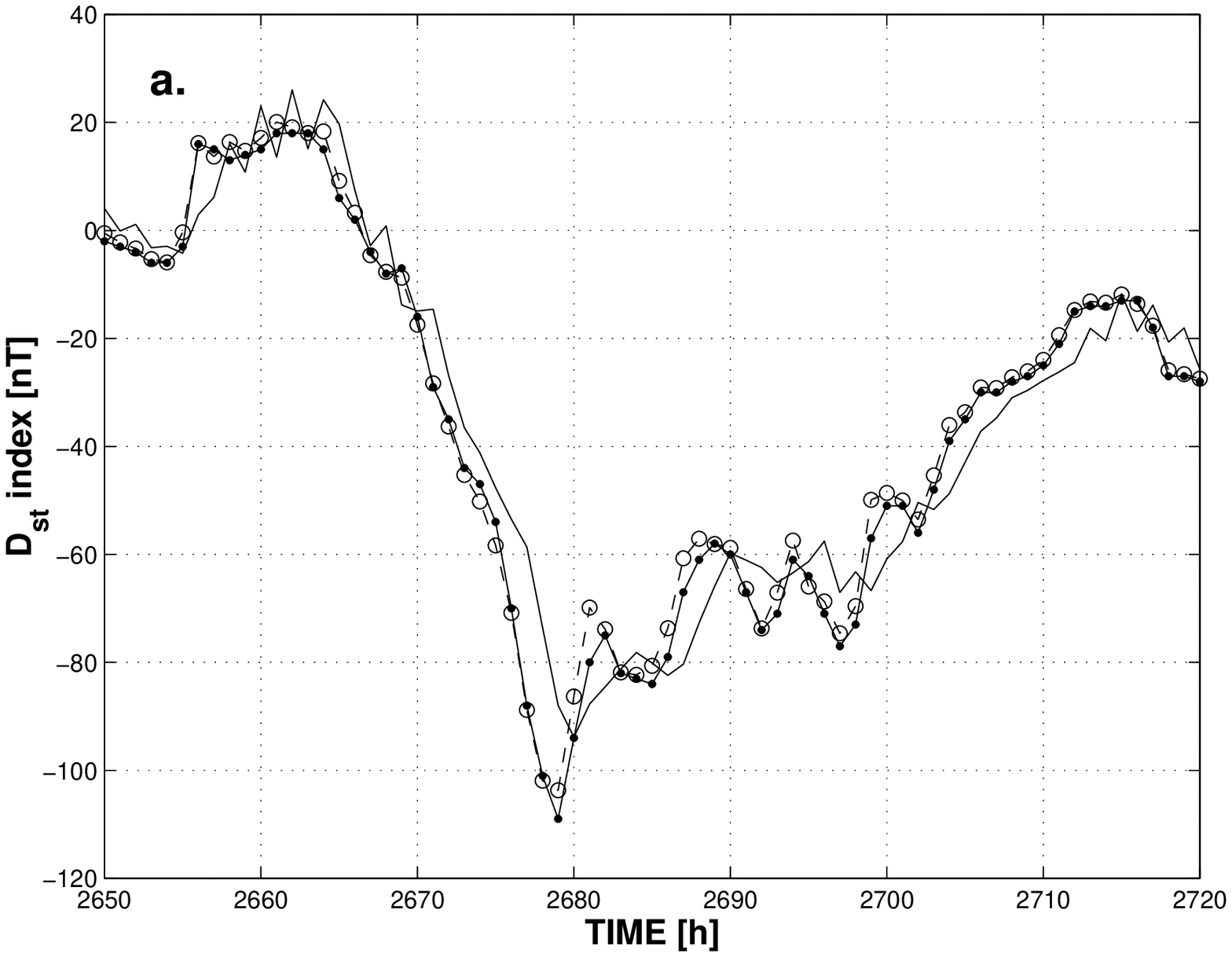}
\includegraphics[width=3.2in]{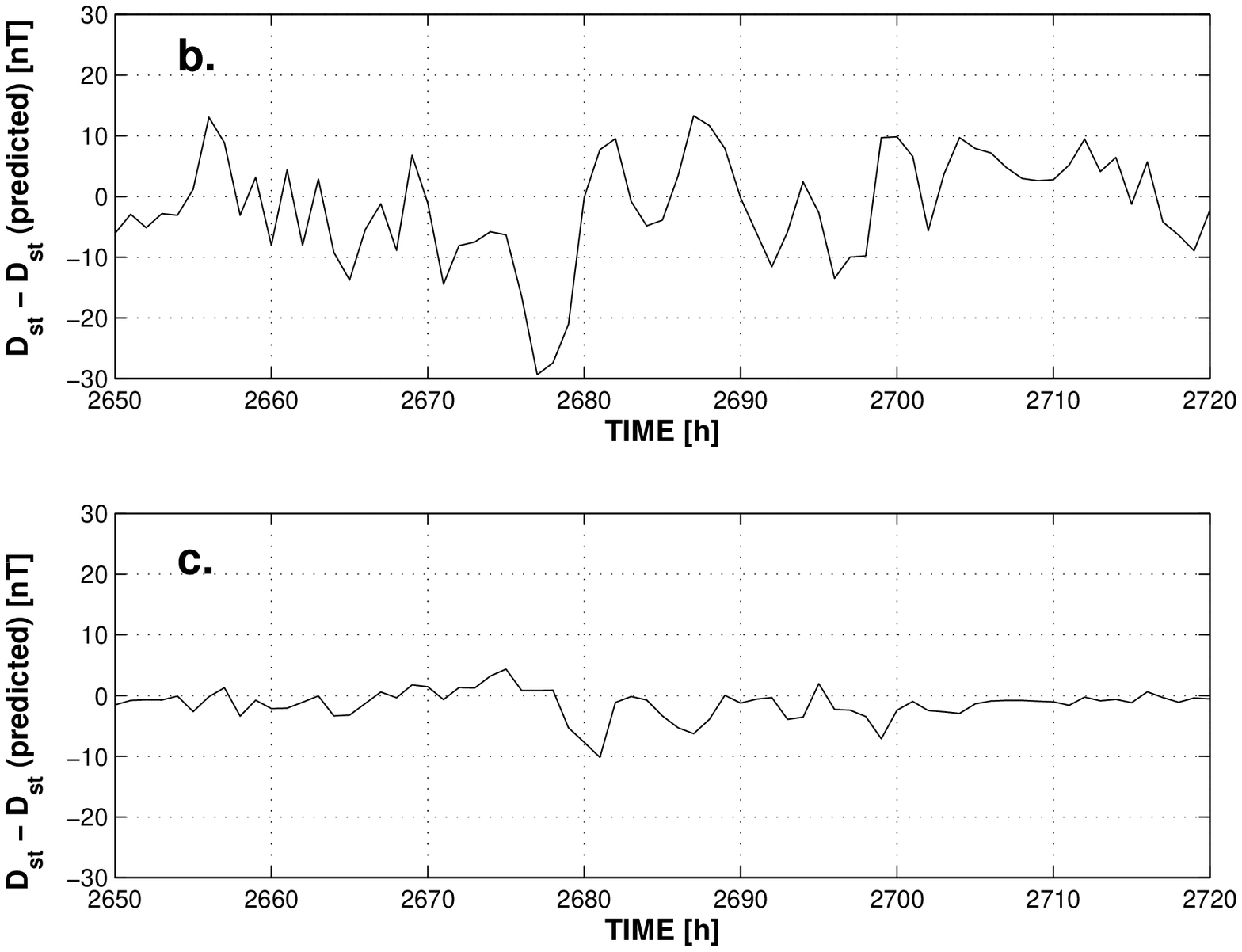}
}
\caption{{\bf a}. 1 hour ahead prediction of $D_{st}$ index for  
           period from March 10, 2001 10:00 UT to March 13, 2001 08:00 UT 
	   ($\bullet$ - actual output; -- - prediction without 
	   H\"{o}lder exponents;  
	   o - prediction with H\"{o}lder exponents of $Pc1$, $Pc2$ and  $D_{st}$ on input;
	   {\bf b}. differences between actual $D_{st}$ and predicted  $D_{st}$ 
	   without H\"{o}lder exponent time series;	
	   {\bf c}. differences between actual $D_{st}$ and predicted $D_{st}$ with 
	   H\"{o}lder exponent time series  on input.}
\end{figure}

\end{document}